\documentclass{epl}

\title{Deviations from the mean field predictions for the phase
behaviour of random copolymers melts}
\shorttitle{ Phase behaviour of random copolymer melts }
\author{J. Houdayer\inst{1,2} \and M. M{\"u}ller\inst{1}}
\institute{
\inst{1} Institut f{\"u}r Physik, Johannes Gutenberg Universit{\"a}t,
D--55099 Mainz, Germany\\
\inst{2} Max Planck Institut f{\"u}r Polymerforschung,
D--55021 Mainz, Germany
}


\begin{document}

\maketitle

\begin{abstract}
We investigate the phase behaviour of random copolymers melts via
large scale Monte Carlo simulations. We observe macrophase separation
into A and B--rich phases as predicted by mean field theory only for
systems with a very large correlation $\lambda$ of blocks along the
polymer chains, far away from the Lifshitz point. For smaller values
of $\lambda$, we find that a locally segregated, disordered
microemulsion--like structure gradually forms as the temperature
decreases. As we increase the number of blocks in the polymers, the
region of macrophase separation further shrinks. The results of our
Monte Carlo simulation are in agreement with a Ginzburg criterium,
which suggests that mean field theory becomes {\em worse} as the
number of blocks in polymers increases.
\end{abstract}

\section{Introduction}
In polymeric systems one often encounters phase equilibria in mixtures
with a large number of components. For instance, polydispersity (i.e.
a continuous distribution of chain lengths) alters the phase
equilibrium between a dense polymer phase and its vapour. Random AB
copolymers constitute another example of such mixtures, each sequence
of A's and B's represents a different component. The phase behaviour
of random copolymers has attracted abiding interest, because these
systems are produced commercially in large amounts and exhibit a rich
phase behaviour. Mean field calculations \cite{SG1,FML} predict a
disordered phase, a macrophase separated phase and a disordered
microphase separated phase, as a function of the incompatibility
$\chi$ of the constituents A and B and the correlation $\lambda$ along
the polymers.

We present the first Monte Carlo study of the phase behaviour of
random copolymers melts and compare our results to the mean field
predictions. For rather short chain lengths we observe all three
phases predicted by the mean field theory, but the region of stability
of the macrophase separation is much smaller than predicted.
Increasing the chain length, we reduce the region of macrophase
separation and increase the deviation from the mean field
prediction. Our simulation results indicate that random copolymer
melts are one of the very few dense polymer systems for which mean
field predictions do not become accurate, even if the chains strongly
interdigitate and interact with many neighbours. This finding is
corroborated by the Ginzburg criterium~\cite{GINZ}.

The layout of our paper is as follows: In the next section we briefly
describe the model and summarise the mean field results pertinent to
our computer simulation study. We give some details about the
simulation technique \cite{JEROME1,JEROME3} in the following
section. The dependence of the phase diagram on the chain length and
correlation length along the chain are discussed and our findings are
rationalised in terms of a Ginzburg criterium. The manuscript closes
with a discussion and an outlook.

\section{Model and Mean Field Predictions}
We consider a dense mixture of linear random copolymers. All chains
consist of a random sequence of $Q$ blocks, each of which comprises
either $M$ monomers of type A or $M$ monomers of type B. The sequence
of $Q$ blocks is characterised by the average composition $f$
(fraction of A--monomers) and the correlation parameter $\lambda$. The
sequences of the polymers are once built by a random polymerisation
process and are then kept fixed. The probability $P_{\rm AB}$ for an
A--block to be followed by a B--block is $P_{\rm AB}=(1-f)(1-\lambda)$
and similarly $P_{\rm BA}=f(1-\lambda)$ (of course, $P_{\rm
AA}=1-P_{\rm AB}$ and $P_{\rm BB}=1-P_{\rm BA}$). For $\lambda=1$,
blocks are completely correlated and the model describes a binary
mixture of homopolymers. For $\lambda=-1$ the polymers consist of a
strictly alternating sequence of A and B--blocks. In the following, we
only consider the symmetric case $f=1/2$ and we denote the polymer
length by $N=QM$.

A and B--monomers repel each other, and the strength of the
repulsion is parameterised by the Flory--Huggins parameter $\chi$.
Upon increasing $\chi$, monomers of the same type segregate to lower
the energy, and a rich phase behaviour is anticipated: Within the mean
field approximation, Fredrickson, Milner, and Leibler~\cite{FML} have
calculated the free energy as a functional of the local composition
$m({\bf r})=\phi_A({\bf r})-f$. For symmetric composition $f=1/2$ it
takes the form:
\begin{equation}
\frac{{\cal F}[m]}{k_{\rm B}T\rho}=\frac{1}{V}\int {\rm d}{\bf r}\;\left\{ 
(\chi_c-\chi)m^2+\frac{4Q}{M\Lambda^2}m^4 +\chi_c \Gamma R_M^2(\nabla m)^2 \right\},
\label{eqn:F}
\end{equation}
with $\Lambda=(1+\lambda)/(1-\lambda)$ and
$\Gamma=(\lambda-\lambda_L)(\lambda+2+\sqrt{3})/[3(1-\lambda^2)]$ for
$\lambda^Q\ll 1$~\cite{COMMENT}. $R_M$ is the radius of gyration of a
block. We assume $\lambda>\lambda_L=-2+\sqrt{3}\approx
-0.268$. Otherwise the coefficient of the square gradient term
becomes negative and higher derivatives have to be considered.

At large incompatibility $\chi > \chi_c=2/(M\Lambda)$, there is
macrophase separation. Since the typical excess of A--monomers along
a random copolymer is of order $M/\sqrt{Q}$ the difference of the
composition of the two phases also is only proportional to
$1/\sqrt{Q}$ and phase separation reduces the energy of a single
molecule by $\chi M$ independently of the number $Q$ of blocks per
chain.

If $\chi$ is increased further, mean field theory predicts a second
transition at $\chi_m$ where the two coexisting phases remix and form
a microphase with no long ranged order. The characteristic length
scale is set by the size $R_M$ of a block and becomes smaller as we
increase $\chi$. The stability region of the macrophase separation
decreases with the number of blocks; $\Delta t \equiv [\chi_m - \chi_c
]/\chi_c$ is of the order $1/Q$.

It has been suggested that the role of fluctuations on the macrophase
transition is similar to that of a binary mixture of homopolymers
\cite{FM1,FML}. In the homopolymer blend, there is only a very narrow
non--classical region about the critical line that is inversely
proportional to the chain length. For smaller correlations, stronger
fluctuation effects are anticipated. In particularly, the isotropic
Lifshitz point at $\lambda=\lambda_L$ does not survive
fluctuations~\cite{DL} and fluctuations will change the order of the
transition towards the microphase separated state from second to first
\cite{SGS,GSS,DE,D2}.

\section{Simulation Technique}
In the Monte Carlo simulations we employ the bond fluctuation
model~\cite{BFM}. Monomers are represented by a cube on a 3D lattice
and block its 8 corners from further occupancy. Monomers along a
polymer are connected via bond vectors of length 2, $\sqrt{3}$,
$\sqrt{5}$, 3 and $\sqrt{10}$. Interactions between monomers are
modelled by a square well potential, which extends over the nearest 54
lattice sites. Monomers of different types repel each other with an
energy $\varepsilon_0$~\cite{COMMENT2} and we define $\varepsilon =
\varepsilon_0/k_{\rm B}T$. This is related to $\chi$ through
$\chi\simeq z\varepsilon$ where $z$ is the number of intermolecular
contacts. This model is an ideal testing bed for comparing simulations
to the predictions above. The phase behaviour~\cite{BLENDS} and
interface properties~\cite{WSMB} of homopolymer mixtures and diblock
copolymers~\cite{MMCOP} have been studied in the framework of this
coarse grained model, and the simulation results have been compared to
mean field predictions of the Gaussian chain model without any
adjustable parameter.

In fact, Swift and Olvera de La Cruz~\cite{MC} have used this model to
investigate the behaviour of random copolymers. Using a local hopping
algorithm, they found evidence for microphase separation at
$\lambda=0$ and a substantially decrease of the centre of mass
diffusion at low temperatures. The slow relaxation corroborates a
frozen--in, glassy structure predicted for low temperatures
\cite{D,GSG2,SS}. Extracting the phase behaviour and its dependence on
the chain length poses a challenge for computer simulations: (i) One
needs an efficient update of the chain conformations on the lattice to
reduce the protracted long time scales associated with the (almost)
frozen structure. (ii) The canonical ensemble is not well suited to
investigate macrophase separation, because the two coexisting phases
(and the interfaces separating them) are simultaneously present in the
simulation cell. Accurately extracting the phase boundaries from
simulations in the canonical ensemble would require inaccessibly large
simulation cells since we expect weak segregation and wide
interfaces. (iii) Due to the finite size of the system, the actual
mixture produced by the random polymerisation process deviates from
the mean composition defined by $f$ and $\lambda$. One should {\it a
priori} average over different realisations of the disorder,
considerably increasing the CPU time required. Due to these
computational difficulties, the way how fluctuations modify the phase
behaviour predicted by mean field calculations could not be addressed
in computer simulations.

To cope with the freezing of the structure, we employ a recently
devised ``wormhole'' algorithm~\cite{JEROME1}, which generalises and
improves the slithering snake algorithm for random copolymers.
Additionally, we use parallel tempering~\cite{HN} to help the
structure equilibrate at low temperatures (large $\chi$). Generally, a
grand-canonical simulation of a random copolymer mixture is not
straightforward, because the mixture contains a large number of
components (i.e. different sequences). The density of the species is
set by the disorder, but their (temperature dependent) chemical
potentials are unknown. If we restrict ourselves, however, to
symmetrical composition $f=1/2$, the system is invariant under
exchanging $\mbox{A}\rightleftharpoons\mbox{B}$ monomers. Hence, it is
possible to switch the identity of all monomers along one chain (and
accept this Monte Carlo move according the Metropolis criterium). This
moves relaxes the composition of the system, which is the order
parameter of the macrophase separation. Owing to this symmetry, the
system is either symmetric (disordered or microphase separated state)
or it coexists with its symmetric in the two phase region. Standard
finite size scaling techniques of grand-canonical simulations can be
applied to accurately locate the phase boundaries.  Finally to avoid
the average over a large number of realisations of the disorder
(i.e. the sequences of the polymers), we devised a method to produce a
``generic'' random copolymer mixture which does not present random
fluctuations. More details on the simulation technique will be
published elsewhere~\cite{JEROME3}.

\section{Results}

\begin{figure}
\resizebox{0.45\linewidth}{!}{\includegraphics{m_curves.eps}}
\hfill
\resizebox{0.4\linewidth}{0.4\linewidth}{\includegraphics{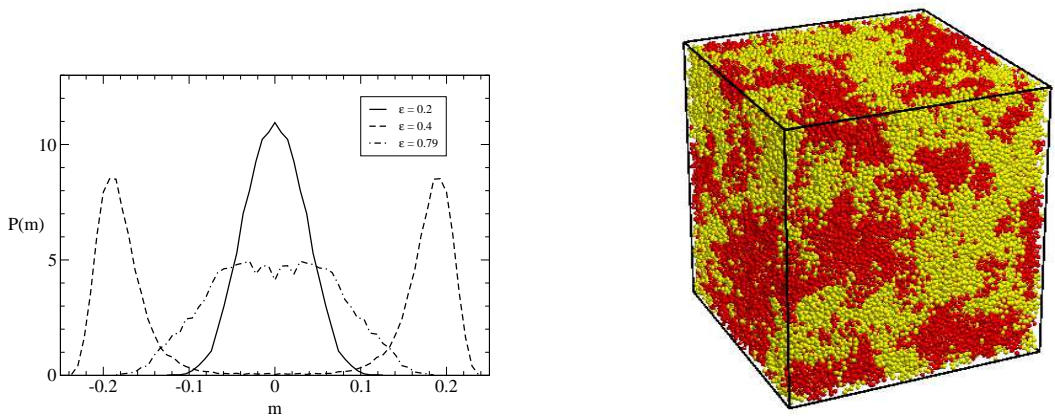}}
\caption{
{\it Left:} Distributions of $m$ (the composition of the system), for
different values of the incompatibility $\varepsilon$. The data here
corresponds to $M=1$, $Q=20$, $\lambda=0.85$ in simulation box of size
$80^3$. {\it Right:} A snapshot of the same system for
$\varepsilon=0.79$ (in a $120^3$ box). The monomers are represented by
small spheres and the bonds are not shown. The colors correspond to
the nature of the monomers (A or B). The typical end-to-end distance
for one polymer is here roughly one 13th of the box size.}
\label{fig_m}
\end{figure}

We have simulated different sets of the parameters $Q$ and $M$, for a
large number of values of $\varepsilon$ and $\lambda$. We found the
following: at small $\varepsilon$ (i.e. high temperature) the system
is homogeneous. Provided that $\lambda$ is large enough, the system
phase separates when one increases the incompatibility (i.e. decrease
the temperature). Finally, at an even larger $\varepsilon$, the system
remixes. This is determined by looking at the distribution $P(m)$ of
the composition (Fig.~\ref{fig_m}). The macrophase separation is a
standard Ising-like second order phase transition and we accurately
determine the transition temperature by mapping $P(m)$ onto the known
distribution of the 3d Ising universality class. From a numerical
point of view, the nature of the remixing transition is unclear, it
could be a simple second order phase transition with strong finite
size effects or there could be a region where more than two phases
coexist. Hence, error bars in the temperature of remixing are much
larger than for the unmixing at higher temperature. The snapshot in
Fig.~\ref{fig_m} shows a typical configuration in the low temperature
phase. Large domains can be observed which have a typical length scale
several times larger than the end-to-end distance of one polymer. On
the other hand, if $\lambda$ is small, we do not see any phase
transition. The distribution $P(m)$ simply broadens and narrows again
as the temperature decreases.

\begin{figure}
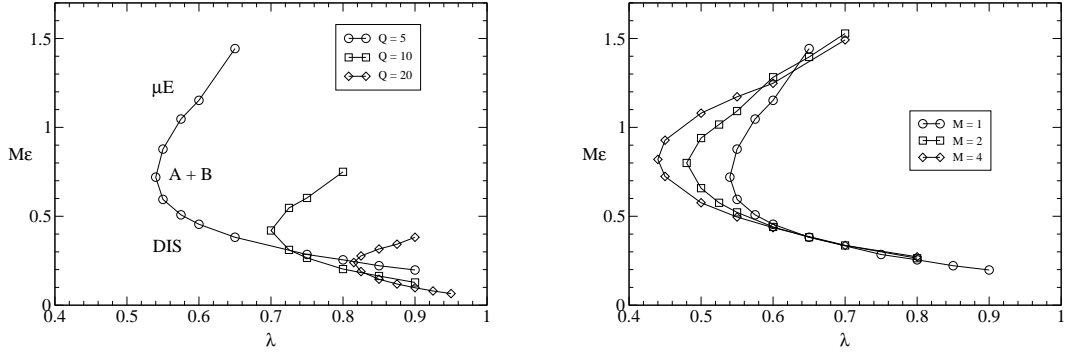

\centerline{
\resizebox{0.45\linewidth}{!}{\includegraphics{phase_Q.eps}}
\hfill
\resizebox{0.45\linewidth}{!}{\includegraphics{phase_M.eps}}
}
\caption{
{\it Left:} Phase boundaries as a function of the correlation
parameter $\lambda$ and the incompatibility $\varepsilon$ for
different values of $Q$ at constant $M=1$. The disordered phase (DIS),
the macrophase separated phase (A+B) and the microemulsion-like phase
($\mu$E) are respectivelly below, on the right and above the
curves. {\it Right:} The same for different values of $M$ at constant
$Q=5$.}
\label{fig_MQ}
\end{figure}

We determined the phase boundaries for different values of $M$ and
$Q$. The topology of the phase diagram we found is similar to the one
of a ternary mixture of two homopolymers and a diblock copolymer (it
is actually the special case $Q=2$ and large $M$). The results are
shown in Fig.~\ref{fig_MQ} using the scaling of mean field theory. On
the left part of Fig.~\ref{fig_MQ} (varying $Q$ at constant $M$), we
see that the phase boundaries do not seem to scale as expected by mean
field. The scaling in $\varepsilon$ does not apply and as one
increases $Q$ the minimal value of $lambda$ for which we observe a
phase separation goes to larger values of $\lambda$ instead of going
to $\lambda_0$.

In view of the unexpectedly large deviations between our simulations
and the mean field predictions, we assess the validity of the latter
approach via the Ginzburg criterium. The mean field approximation is
accurate, if fluctuations $\langle m^2 \rangle$ of the composition in
a volume of the size of the correlation length $\xi$ are small
compared to the difference of composition between the two coexisting
phases. Using the mean field functional (equation~\ref{eqn:F}) we
calculate the composition of the two coexisting phases:
$m=\pm\sqrt{\Lambda |t|/4Q}$ where $t=(\chi_c-\chi)/\chi_c$ denote the
distance to the onset of macrophase separation. The correlation length
above the critical temperature is given by
$\xi^+=R_M\sqrt{\Gamma/|t|}/2$ and the critical amplitude is reduced
by a factor $\sqrt{2}$ below the critical point. The scale of the
correlation length is set by the size of an individual block $R_M$ and
does not depend on the number $Q$ of blocks. The susceptibility above
the critical point takes the form
$V\langle m^2\rangle^+=M\Lambda/[4\rho|t|]$ and the critical
amplitude is a factor $2$ lower in the macrophase separated
state. Using these expressions, we obtain for the Ginzburg criterium:
\begin{equation}
t \gg {\rm Gi}=\frac{16}{\Gamma^3} Q^2 \left( \frac{M}{\rho R_M^3}
\right)^2 \sim Q^2\frac1M.
\end{equation}
The last factor measures the inverse degree of interdigitation of
blocks. This factor resembles the behaviour of binary homopolymer
blends, except for the important difference that it is only
proportional to the inverse length of a block ($1/M$) and not to that
of the whole chain ($1/MQ$). In marked contrast to the homopolymer
blend, the Ginzburg number increases and the region of validity of the
mean field theory decreases as we increase the number $Q$ of
blocks. Our finding disagrees with previous estimates which assumed a
behaviour similar to a binary mixture of homopolymers.

We expect fluctuations to shift the macrophase separation to larger
values of the incompatibility by an amount $\delta t \sim QM/\rho
R_M^3$. For long random copolymers (i.e. large $Q$) this shift exceeds
by far the stability range of the macrophase separated state which is,
in the framework of the mean field calculation, of order $\Delta t
\sim 1/Q\ll\delta t$. Of course, fluctuations also affect the onset of
microphase separation and shift it to higher values of $\varepsilon$. Our
simulations, however, indicate that the stability region of the
macrophase separation decreases with increasing number of blocks $Q$.

On the other hand, the large $M$ behaviour (at constant $Q$)
approaches the mean field predictions as can be seen on the right part
of Fig.~\ref{fig_MQ}. This is consistent with the fact that
$G_i\rightarrow 0$ as $M\rightarrow\infty$ which indicates that mean
field should be correct in this limit. This limit is not generic for
copolymers built by random polymerisation of blocks because the size
$M$ of the blocks would have to increase with increasing number of
blocks. Nevertheless the case $Q=2$ corresponds to a ternary mixture
of two homopolymers and a diblock copolymer in which case the
$M\rightarrow\infty$ is the relevent limit and is known to approach
mean field~\cite{HS}.

\section{Summary}

Using extensive Monte Carlo simulations, we have studied the phase
diagram of random copolymer melts. The region where macrophase
separation occurs is much smaller than predicted by mean field, and
this region further shrinks as the number $Q$ of blocks in the
polymers increases. We can then conclude for this system that
surprisingly, mean field becomes worse as the size of the polymers
increases. On the other hand, the mean field behaviour is recovered in 
the limit of large blocks ($M\rightarrow\infty$). Both observations
are corroborated by the Ginzburg criterium.

The unexpected observation, that mean field theory fails to describe
the qualitative behaviour in the limit of infinitely long chains, is
traced back to the weak segregation and the small correlation length
due to the disorder in the polymer sequences.

Future work should address three subjects not treated here: (i) The
nature of the low temperature phase, in particular its spacial
structure. (ii) The type of the transition between the macrophase
separated region and the low temperature phase. We are currently
investigating the possible existence of a 3 phase region, using grand
canonical simulations with chemical potentials (at small $Q$ to reduce
the number of different chemical species). (iii) The scaling of
$\varepsilon$ in the large $Q$ regime.

\acknowledgments
It is a pleasure to thank K. Binder and A. Yethiraj for helpful and
enjoyable discussions. J.H. thanks the Max Planck Society for a
fellowship and M.M. was supported by a Heisenberg--stipend. Generous
allocations of computer time at the NIC J{\"u}lich are gratefully
acknowledged.


\begin{thebibliography}{99}
\bibitem{COMMENT}
In the limit $\lambda \to 1$ we obtain the expansion for a binary
homopolymer blend with $\Lambda \sim \Gamma \sim Q$.
\bibitem{COMMENT2}
There is no attraction between monomers of the same type to avoid
large fluctuations of the total density, a liquid--vapour phase
separation, or even crystallisation. Note that, according to mean
field, the energy scale of interest is of order $k_{\rm B}T/M$ per monomer
and not $k_{\rm B}T/N$ as usual.
\end{thebibliography}
\end{document}